\documentstyle[12pt,fleqn]{article}
\input epsf

\newcommand{\la}{\langle}
\newcommand{\ra}{\rangle}

\newcommand{\pa}{\partial}
\renewcommand{\a}{\alpha}
\renewcommand{\b}{\beta}
\renewcommand{\c}{\gamma}

\newcommand{\e}{\epsilon}
\newcommand{\m}{\mu}
\newcommand{\n}{\nu}

\begin{document}

\rightline{THES-TP 95/9}
\vspace{1cm}
\begin{center}
\large
{\bf{$C_{T}$ and $C_{J}$ up to next-to-leading order in
$1/N$ }}\\{\bf{  in the
Conformally Invariant $O(N)$ Vector Model for $2<d<4$}}
\vspace{2cm}
\normalsize
\\
{\bf{Anastasios C. Petkou}}\footnote{e-mail: vlachos@athena.auth.gr} \\
Department of Theoretical Physics \\ Aristotle University of
Thessaloniki\\ Thessaloniki 54006,  Greece
\end{center}

\vspace{2cm}
\begin{abstract}
Using Operator Product Expansions and a graphical ansatz for the
four-point function of the fundamental field $\phi^{\alpha}(x)$ in the
conformally invariant $O(N)$ vector model, we calculate the
next-to-leading order in $1/N$
values  of  the quantities $C_{T}$ and $C_{J}$. We check the
results  against  what is
expected from possible generalisations of the $C$- and $k$-theorems in
higher dimensions and also against known three-loop calculations in a $O(N)$
invariant $\phi^{4}$ theory for  $d=4-\epsilon$.
\end{abstract}
\vfill

\section{Introduction}

Zamolodchikov's $C$-theorem \cite{Zamolodchikov1} states that  there
exist  a
quantity which is
monotonously decreasing along the renormalisation group (RG) flow from
the UV to the IR fixed points of unitary
two-dimensional quantum field theories (QFT's). At the fixed points, which
correspond to conformal field theories (CFT's) \cite{Schroer}, this
coincides with the Virasoro central charge and the conformal
anomaly. Consequently, it has been suggested that a possible
generalisation of the
$C$-theorem in higher
dimensions involves the conformal anomaly
\cite{Cardy1}. This is a promising yet not  well worked out
idea, (see \cite{anastasios1} and
references therein), since there does not seem
to exist i.e. in four
dimensions explicit CFT models  other than the trivial bosonic,
fermionic and spin-1
theories \footnote{See however \cite{Seiberg}}.

An alternative definition for the two-dimensional central charge is through
the coefficient $C_{T}$ of the two-point function of the energy
momentum  tensor
$T_{\m\n}(x)$. The latter has a unique form in any dimension $d$
\cite{anastasios1}, therefore
it is conceivable that $C_{T}$ is related to the fixed point
value of a possible generalisation of the $C$-function in three
dimensions which is relevant for studies of statistical mechanical
systems. Conveniently, in three
dimensions there is  a number of non-trivial
CFT's \cite{Gracey} and such an idea can be explicitly checked out.

In this
letter we briefly report and discuss some of the
results obtained in
\cite{anastasios2} for  the Euclidean conformally invariant $O(N)$
vector model in
$2<d<4$ (the physically relevant  dimension is
  $d=3$). This model
provides an  example of  a RG flow from an UV fixed point
(free massless theory) to an IR one (non-trivial CFT) which can be
explicitly studied. In
\cite{anastasios2} we calculated, among others, the value of
$C_{T}$ at the non-trivial IR fixed point up to
next-to-leading order in $1/N$ and we found it  smaller that the value
of $C_{T}$ at the UV fixed point in accordance with a possible
generalisation of the $C$-theorem in higher dimensions.

Recently, the $k$-theorem \cite{Cardona} was proved for
two-dimensional QFT's stating that the Kac-Moody algebra level is
the fixed point value of a
quantity which is monotonously decreasing along the RG flow from the
UV to the IR. A possible generalisation in higher dimensions of the
Kac-Moody algebra level is through the coefficient $C_{J}$ of the
two-point function of an internal symmetry conserved current
$J_{\mu}(x)$. The latter has a unique form in any dimension $d$
\cite{anastasios1}.  In \cite{anastasios2} we also calculated the value of
$C_{J}$
up to next-to-leading order in $1/N$ at the IR fixed
point of the $O(N)$ vector model and we found it smaller than the value
of  $C_{J}$ at
the UV fixed point in accordance with a possible generalisation of the
$k$-theorem.

\section{General Framework for Calculations in the Conformally
Invariant $O(N)$  Vector Model }

The starting point of our calculations is the four-point function of
the  fundamental field \cite{Zamolodchikov2}    $\phi^{\a}(x)$,
$\a=1,2,..,N$  in the
$O(N)$ vector model. This, being dependent from conformal
invariance on two
variables \cite{Ginsparg} , can be written as \cite{anastasios2}
\begin{eqnarray}
\Phi^{\a\b\c\delta}(x_{1},x_{2},x_{3},x_{4})& = &
H(\eta,x)\,F_{S}(u,v)\,\delta^{\a\b}\delta^{\c\delta} \nonumber \\
 & &
{}\hspace{-2.5cm}+H(\eta,x)\,F_{V}(u,v)\,\frac{1}{2}(\delta^{\a\c}\delta^{
\b\delta}
-\delta^{\a\delta}\delta^{\b\c})
\nonumber \\
 &  &
{}\hspace{-2.5cm}+H(\eta,x)\,F_{T}(u,v)\,\frac{1}{2}(\delta^{\a\c}\delta^{
\b\delta}
+\delta^{\a\delta}\delta^{\b\c}-\frac{2}{N}\delta^{\a\b}\delta^{\c\delta})
,
\end{eqnarray}
where
\begin{eqnarray}
F_{S}(u,v) & = & F(u,v)
+\frac{1}{N}\left(F(\frac{1}{u},\frac{v}{u})+F(\frac{u}{v},\frac{1}{v})
\right),
\nonumber \\
F_{V}(u,v) & = & F(\frac{1}{u},\frac{v}{u})-F(\frac{u}{v},\frac{1}{v}),
\,\,\,\,\,F_{T}(u,v) =
F(\frac{1}{u},\frac{v}{u})+F(\frac{u}{v},\frac{1}{v}), \\
H(\eta,x) & = & \frac{1}{(x_{12}^{2}x_{34}^{2}x_{13}^{2}x_{24}^{2}x_{14}^{2}
x_{23}^{2})^{\frac{1}{3}\eta}},
\end{eqnarray}
and $F(u,v)=F(v,u)$ is an arbitrary function of the two invariant ratios
\begin{equation}
u=\frac{x_{12}^{2}x_{34}^{2}}{x_{13}^{2}x_{24}^{2}}
\,\,\,\,\,\,\mbox{and}\,\,\,\,\,
v=\frac{x_{12}^{2}x_{34}^{2}}{x_{14}^{2}x_{23}^{2}}.
\end{equation}

Next, we make the basic assumption that the field algebra of a $O(N)$
invariant CFT is qualitatively similar to the field algebra  of a
free
theory of $N$ massless scalar fields. Therefore, we write for the OPE of
$\phi^{\a}(x)$ with itself
\begin{eqnarray}
\phi^{\a}(x_{1})\phi^{\b}(x_{2})  & = &
C_{\phi}\frac{1}{x_{12}^{2\eta}}\delta^{\a\b}
+C^{\phi\phi O}(x_{12},\pa_{2})\,O(x_{2})\delta^{\a\b}
\nonumber \\
 &   & {}+\frac{g_{\phi\phi
J}}{C_{J}}\,\frac{(x_{12})_{\mu}}{(x_{12}^{2})^{\eta -\mu
+1}}\,J_{\mu}^{\a\b}(x_{2}) +\cdots \nonumber \\
&  & {}-\frac{g_{\phi\phi
T}}{C_{T}}\,\frac{(x_{12})_{\mu}(x_{12})_{\nu}}{(x_{12}^{2})^{\eta
-\mu +1}}\,T_{\mu\nu}(x_{2})\,\delta^{\a\b}+\cdots ,
\end{eqnarray}
with $\mu=d/2$. Namely, the most singular terms \footnote{Other
possible  fields
neglected in  (5) include all symmetric traceless rank-2 $O(N)$ tensors.}
as $x_{12}^{2}\rightarrow 0$ in the OPE
(5) are assumed to be, apart from the contribution of the
unit field,
 the  coefficients of the
$O(N)$ conserved vector current $J_{\mu}^{\a\b}(x)$, of the
(traceless)  energy
momentum tensor $T_{\mu\nu}(x)$ and also of
some   scalar field $O(x)$ with
dimension $\eta_{o}$,  $0< \eta_{o}<d$, whose two-point function
in normalised as
\begin{equation}
\la O(x_{1})\, O(x_{2})\ra =
C_{O}\frac{1}{x_{12}^{2\eta_{o}}}.
\end{equation}
$C_{\phi}$ is the normalisation of the two-point function of
$\phi^{\a}(x)$. The couplings $g_{\phi\phi J}$ and $g_{\phi\phi T}$ of
the three-point
functions $\la\phi\phi J\ra$ and $\la\phi\phi T\ra$ respectively can
be found  \cite{anastasios2} (see also \cite{Cardy2}) from the Ward
identities  to be
\begin{equation}
g_{\phi\phi J}=\frac{1}{S_{d}}C_{\phi},\,\,\,\,\,\,\,\, g_{\phi\phi
T}= \frac{d\eta}{(d-1)S_{d}}C_{\phi},\,\,\,\,\,
S_{d}=2\pi^{\frac{1}{2}d}/ \Gamma({\textstyle{\frac{1}{2}}}d).
\end{equation}
The coupling $g_{\phi\phi O}$ of the three-point function $\la\phi\phi
O\ra$ and the field dimensions $\eta$, $\eta_{o}$ are  dynamical
parameters of the theory. The full OPE coefficient $C^{\phi\phi
O}(x_{12},\pa_{2})$ can be found in a closed form using conformal
integration techniques \cite{anastasios2}. Substituting the OPE (5)
into  the
four-point function (1) we  obtain
the most singular terms of the latter in the limit as $x_{12}^{2}$,
$x_{34}^{2}\rightarrow 0$ {\it{independently}}. For completeness we
give the form for the conformally invariant two-point functions of
$T_{\m\n}(x)$ and $J_{\m}^{\a\b}(x)$ as \cite{anastasios1}

\begin{eqnarray}
\la T_{\mu\nu}(x_{1})T_{\rho\sigma}(x_{2})\ra &  = &
C_{T}\frac{I_{\mu\nu ,\rho\sigma}(x_{12})}{x_{12}^{2d}}, \\
\la J_{\mu}^{\a\b}(x_{1})J_{\nu}^{\c\delta}(x_{2})\ra & = &
C_{J}\frac{I_{\mu\nu}(x_{12})}{x_{12}^{2(d-1)}}\,(\delta^{\a\c}
\delta^{\b\delta}-\delta^{\a\delta}\delta^{\b\c}),
\end{eqnarray}
with
\begin{eqnarray}
I_{\mu\nu ,\rho\sigma}(x) & = & \frac{1}{2}\Bigl(
I_{\mu\rho}(x)I_{\nu\sigma}(x)+I_{\mu\sigma}(x)I_{\nu\rho}(x)\Bigl)
{}-{}\frac{1}{d}\delta_{\mu\nu}\delta_{\rho\sigma}, \\
I_{\mu\nu}(x) & = &
\delta_{\mu\nu}-2\frac{x_{\mu}x_{\nu}}{x^{2}}.
\end{eqnarray}
After some algebra whose details are explained in \cite{anastasios2}
and using (8) and (9) above,
the leading terms of $F_{S}(u,v)$ and $F_{V}(u,v)$ in (2) as
$x_{12}^{2}$, $x_{34}^{2}\rightarrow 0$ are found to be
\begin{eqnarray}
F_{S}(u,v) & \equiv & {\cal{F}}_{S}(Y,W) \nonumber \\
  =  C_{\phi}^{2}W^{-\frac{2}{3}\eta} \!\!\!\!\!\!\!\!\!\!\!\!\!\!& &
+\frac{g_{\phi\phi
O}^{2}}{C_{O}}\,W^{\frac{1}{2}\eta_{o}-\frac{2}{3}\eta}\,
\Bigg(1-\frac{\eta_{o}^{2}}{32(\eta_{o}
+1)}Y^{2}+\frac{\eta_{o}^{3}}{16(\eta_{o}
+1)(\eta_{o}+1-\mu)}\,W\,\Bigg)\nonumber \\
&   & {}+\frac{g_{\phi\phi T}^{2}}{C_{T}}\,W^{\mu
-1-\frac{2}{3}\eta}\,\Bigg(\frac{1}{4}\,Y^{2}-\frac{1}{d}\,W\Bigg){}
+\cdots,
 \\
F_{V}(u,v) & \equiv &  {\cal{F}}_{V}(Y,W) =  \frac{1}{2}\frac{g_{\phi\phi
J}^{2}}{C_{J}}\,W^{\mu -1-\frac{2}{3}\eta}\,Y+\cdots.
\end{eqnarray}
We have used for convenience the two new independent variables
\begin{eqnarray}
Y=1-\frac{v}{u} & = & 2\frac{1}{x_{24}^{2}}\left[(x_{12}\cdot
x_{34})-2\frac{(x_{12}\cdot x_{24})(x_{34}\cdot
x_{24})}{x_{24}^{2}}\right]+\cdots, \\
W=(uv)^{\frac{1}{2}} & = &
\frac{x_{12}^{2}x_{34}^{2}}{x_{24}^{4}}+\cdots,
\end{eqnarray}
and the dots in (12), (13) stand for less singular terms in the limit
$Y , W\rightarrow 0$.

It is clear now that having an expression for the
four-point function (1) we can take suitable short distance limits
and compare them  with the above formulae (12) and (13). This would
determine the values of the coupling $g_{\phi\phi O}$, the field
dimensions $\eta$, $\eta_{o}$ and the
wanted quantities $C_{T}$ and $C_{J}$. Such is the case for the UV
fixed point of the $O(N)$ vector model, which corresponds to a free
theory of $N$ massless scalar fields, when the full expression
for the four-point function can be found using Wick's theorem with
elementary contraction the two-point function of $\phi^{\a}(x)$. This
result indicates a graphical  representation for $F_{f}(u,v)$ as shown
in Fig.1, where the solid lines stand for the two-point function of
$\phi^{\a}(x)$ and the subscript $f$ stands for ``free field theory''.
\begin{figure}[t]

\setlength{\unitlength}{0.01100in}%
\begingroup\makeatletter
% extract first six characters in \fmtname
\def\x#1#2#3#4#5#6#7\relax{\def\x{#1#2#3#4#5#6}}%
\expandafter\x\fmtname xxxxxx\relax \def\y{splain}%
\ifx\x\y   % LaTeX or SliTeX?
\gdef\SetFigFont#1#2#3{%
  \ifnum #1<17\tiny\else \ifnum #1<20\small\else
  \ifnum #1<24\normalsize\else \ifnum #1<29\large\else
  \ifnum #1<34\Large\else \ifnum #1<41\LARGE\else
     \huge\fi\fi\fi\fi\fi\fi
  \csname #3\endcsname}%
\else
\gdef\SetFigFont#1#2#3{\begingroup
  \count@#1\relax \ifnum 25<\count@\count@25\fi
  \def\x{\endgroup\@setsize\SetFigFont{#2pt}}%
  \expandafter\x
    \csname \romannumeral\the\count@ pt\expandafter\endcsname
    \csname @\romannumeral\the\count@ pt\endcsname
  \csname #3\endcsname}%
\fi
\endgroup
\begin{picture}(190,71)(-30,530)
\thicklines
\put(320,590){\line( 1, 0){ 60}}
\put(320,545){\line( 1, 0){ 60}}

\put(100,565){\makebox(0,0)[lb]{\smash{\SetFigFont{10}{14.4}{rm}
$F_{f}(u,v)\,\,\,\,\,\,\,\,\,\, =\,\,\,\,\,\,\,\,\,\, H^{-1}(x,\eta)
\,\,\,\,\,\,
\times $}}}
\put(343,608){\makebox(0,0)[lb]{\smash{\SetFigFont{10}{14.4}{rm}$
{\cal{G}}_{0}$}}}

\put(310,595){\makebox(0,0)[lb]{\smash{\SetFigFont{10}{14.4}{rm}$x_{1}$}}}
\put(380,595){\makebox(0,0)[lb]{\smash{\SetFigFont{10}{14.4}{rm}$x_{2}$}}}
\put(310,535){\makebox(0,0)[lb]{\smash{\SetFigFont{10}{14.4}{rm}$x_{3}$}}}
\put(380,535){\makebox(0,0)[lb]{\smash{\SetFigFont{10}{14.4}{rm}$x_{4}$}}}
\end{picture}
\caption{The Graphical Expansion for
$F_{f}(u,v)$}

\end{figure}
{}From $F_{f}(u,v)$ we find $F_{S,f}(u,v)$ and $F_{V,f}(u,v)$ as in (2)
and we take their short distance limits as $Y$, $W\rightarrow 0$
{\it{independently}}. The resulting expressions are then compared with
(12) and (13) and yield the values of the various parameters in
the theory as
\begin{eqnarray}
\eta =\frac{d}{2}-1,\,\,\,\,\,\eta_{o}=d-2,\,\,\,\,\,g_{\phi\phi
O}^{2}=\frac{2}{N}C_{O}\,C_{\phi}^{2}, & &\nonumber \\
C_{T}=N\frac{d}{(d-1)S_{d}^{2}},\,\,\,\,\,
C_{J}=\frac{2}{(d-2)S_{d}^{2}}.
\end{eqnarray}
The values of the various parameters given in (16) are in agreement
with results given e.g. in \cite{anastasios1} for the theory of $N$
massless scalar fields in any dimension $d$.

Next, we propose that a graphical expansion for a non-trivial $F(u,v)$
can be obtained by introducing a conformally invariant vertex into the
theory. This vertex is assumed to describe the interaction of
$\phi^{\a}(x)$ with an arbitrary scalar field ${\tilde{O}}(x)$, which
is a $O(N)$ singlet and has dimension $\tilde{\eta}_{o}$ with
$0<\tilde{\eta}_{o}<d$, whose two-point function we represent as a
dashed line. The three-point function $\la\phi\phi\tilde{O}\ra$ has a
coupling constant $g_{*}$ which has to be determined from the dynamics
of the theory. Moreover, we assume that the amplitudes for $n$-point
functions of $\phi^{\a}(x)$ with $n\geq 4$ in our non-trivial CFT are
constructed in terms of skeleton graphs with no self-energy or vertex
insertions and internal lines corresponding to the two-point
functions of $\phi^{a}(x)$ and $\tilde{O}(x)$. Here we only consider
graphs involving vertices corresponding  to the fully amputated three-point
function $\la\phi\phi\tilde{O}\ra$ which are graphically represented
as dark blobs \footnote{For the interesting case of graphs involving
the vertex formed by
three $\tilde{O}$ fields see \cite{anastasios2}}. Symmetry factors are
determined as in the usual Feynman perturbation expansion
\footnote{Graphical
expansions in field theory are usually connected with a Lagrangian
formalism. In the present work however, we consider a formulation for
a non-trivial CFT based on a graphical expansion without explicit
reference to an underlying Lagrangian.}.
We denote by $F^{\tilde{\eta}_{o}}(u,v)$ the amplitude of interest in
our  graphical treatment of the four-point function (1). The first few
graphs  in the skeleton expansion for this amplitude in increasing
order  according to the number of vertices are displayed in Fig.2.
\begin{figure}[t]

\setlength{\unitlength}{0.01100in}%
\begingroup\makeatletter
% extract first six characters in \fmtname
\def\x#1#2#3#4#5#6#7\relax{\def\x{#1#2#3#4#5#6}}%
\expandafter\x\fmtname xxxxxx\relax \def\y{splain}%
\ifx\x\y   % LaTeX or SliTeX?
\gdef\SetFigFont#1#2#3{%
  \ifnum #1<17\tiny\else \ifnum #1<20\small\else
  \ifnum #1<24\normalsize\else \ifnum #1<29\large\else
  \ifnum #1<34\Large\else \ifnum #1<41\LARGE\else
     \huge\fi\fi\fi\fi\fi\fi
  \csname #3\endcsname}%
\else
\gdef\SetFigFont#1#2#3{\begingroup
  \count@#1\relax \ifnum 25<\count@\count@25\fi
  \def\x{\endgroup\@setsize\SetFigFont{#2pt}}%
  \expandafter\x
    \csname \romannumeral\the\count@ pt\expandafter\endcsname
    \csname @\romannumeral\the\count@ pt\endcsname
  \csname #3\endcsname}%
\fi
\endgroup
\begin{picture}(444,120)(40,570)
\thicklines
\put(165,665){\line( 1, 0){ 60}}
\put(165,615){\line( 1, 0){ 60}}
\put(285,615){\line( 1, 0){ 60}}
\put(405,665){\line( 1, 0){ 60}}
\put(405,615){\line( 1, 0){ 60}}
\put(285,665){\line( 1, 0){ 60}}
\put(315,665){\circle*{8}}
\put(420,665){\circle*{8}}
\put(450,665){\circle*{8}}
\put(315,615){\circle*{8}}
\put(420,615){\circle*{8}}
\put(450,615){\circle*{8}}

\multiput(315,665)(0.00000,-4.00000){13}{\line( 0,-1){  2.000}}
\multiput(420,665)(0.00000,-4.00000){13}{\line( 0,-1){  2.000}}
\multiput(450,665)(0.00000,-4.00000){13}{\line( 0,-1){  2.000}}
\thinlines
\put(405,700){\line(-1, 0){ 15}}
\put(390,700){\line( 0,-1){120}}
\put(390,580){\line( 1, 0){ 15}}
\put(550,700){\line( 1, 0){ 15}}
\put(565,700){\line( 0,-1){120}}
\put(565,580){\line(-1, 0){ 15}}
\thicklines
\put(160,670){\makebox(0,0)[lb]{\smash{\SetFigFont{10}{14.4}{rm}$x_{1}$}}}
\put(220,670){\makebox(0,0)[lb]{\smash{\SetFigFont{10}{14.4}{rm}$x_{2}$}}}
\put(160,605){\makebox(0,0)[lb]{\smash{\SetFigFont{10}{14.4}{rm}$x_{3}$}}}
\put(220,605){\makebox(0,0)[lb]{\smash{\SetFigFont{10}{14.4}{rm}$x_{4}$}}}
\put(280,605){\makebox(0,0)[lb]{\smash{\SetFigFont{10}{14.4}{rm}$x_{3}$}}}
\put(280,670){\makebox(0,0)[lb]{\smash{\SetFigFont{10}{14.4}{rm}$x_{1}$}}}
\put(340,670){\makebox(0,0)[lb]{\smash{\SetFigFont{10}{14.4}{rm}$x_{2}$}}}
\put(340,605){\makebox(0,0)[lb]{\smash{\SetFigFont{10}{14.4}{rm}$x_{4}$}}}
\put(400,670){\makebox(0,0)[lb]{\smash{\SetFigFont{10}{14.4}{rm}$x_{1}$}}}
\put(460,670){\makebox(0,0)[lb]{\smash{\SetFigFont{10}{14.4}{rm}$x_{2}$}}}
\put(400,605){\makebox(0,0)[lb]{\smash{\SetFigFont{10}{14.4}{rm}$x_{3}$}}}
\put(460,605){\makebox(0,0)[lb]{\smash{\SetFigFont{10}{14.4}{rm}$x_{4}$}}}
\put(40,637){\makebox(0,0)[lb]{\smash{\SetFigFont{10}{14.4}{rm}$
H(x,\eta)\,F^{\tilde{\eta}_{o}}(u,v)$ = }}}
\put(250,637){\makebox(0,0)[lb]{\smash{\SetFigFont{10}{14.4}{rm}$+$}}}
\put(350,637){\makebox(0,0)[lb]{\smash{\SetFigFont{10}{14.4}{rm}$+$}}}

\put(475,637){\makebox(0,0)[lb]{\smash{\SetFigFont{10}{14.4}{rm}$+\,\,\,\,\,
(x_{3}\leftrightarrow x_{4}$)}}}
\put(575,637){\makebox(0,0)[lb]{\smash{\SetFigFont{10}{14.4}{rm}$+\cdots$}}}

\end{picture}

\caption{The Skeleton Graph Expansion for
$H(x,\eta)\,F^{\tilde{\eta}_{o}}(u,v)$}

\end{figure}

The crucial consistency requirement regarding the present work is that
amplitudes  constructed according to graphical expansions such as the
one in Fig.2,  correspond to  CFT's having operator content in the
agreement with the OPE ansatz (5) and are therefore compatible with
amplitudes obtained by straightforward application of this ansatz on
$n$-point functions. Without further input at this point we have no
intrinsic means in estimating
the magnitude of
$g_{*}$ and hence we cannot hope to obtain  a weak coupling
expansion.  However, on account of the  $O(N)$ symmetry we
subsequently see that the assumption   $g_{*}^{2}=O(1/N)$ leads
naturally to  a
well defined perturbation expansion in
$1/N$ for the theory. Therefore, from now on we consider the theory
for large $N$.

Details on the calculation of the amplitudes in Fig.2 are given in
\cite{anastasios2}.  Here we just mention that these calculations are
greatly facilitated by the use of
the DEPP formula \cite{DEPP}
\begin{equation}
\int
d^{d}x\frac{1}{(x_{1}-x)^{2a_{1}}(x_{2}-x)^{2a_{2}}(x_{3}-x)^{2a_{3}}}
= \frac{U(a_{1},a_{2},a_{3})}{(x_{12}^{2})^{\mu
-a_{3}}(x_{13}^{2})^{\mu -a_{2}}(x_{23}^{2})^{\mu
-a_{1}}},
\end{equation}
which is valid for
$a_{1}+a_{2}+a_{3}=d$, with
\begin{equation}
U(a_{1},a_{2},a_{3})=\pi^{\mu}\frac{\Gamma(\mu
-a_{1})\Gamma(\mu -a_{2})\Gamma(\mu
-a_{3})}{\Gamma(a_{1})\Gamma(a_{2})\Gamma(a_{3})}.
\end{equation}
In order to compare the
formulae obtained from this calculation with the
algebraically obtained expressions (12) and (13), we find that we need  to
identify the field $O(x)$ in the OPE ansatz (5), {\it{either}} with
$\tilde{O}(x)$
{\it{or}} with the {\it{shadow field}} \footnote{This is a scalar field
with dimension $d-\tilde{\eta}_{o}$. For more on the notion of
{\it{shadow fields}} in CFT see \cite{Ferrara}.} of the latter. The
second possibility leads to a non-unitary theory which may be related
to the free theory of $N$ massless scalars \cite{anastasios2}. Here we
are concerned with the first possibility above when we set
$\eta_{o}=\tilde{\eta}_{o}$ and $g_{\phi\phi O}=g_{*}$. Expanding then
the parameters of the theory in $1/N$ and having as only input that
the leading order value of $\eta=\mu-1$, we obtain after some algebra
\begin{eqnarray}
\eta & = &
\mu-1+\frac{1}{N}\eta_{1},\,\,\,\,\,\,\,\,\eta_{1}  =\frac{2\,
\Gamma(2\mu-2)}{\Gamma(1-\mu)\Gamma(\mu)\Gamma(\mu+1)\Gamma(\mu-2)}\,,\\
\eta_{o} & = &
2+\frac{1}{N}\,\frac{4(2\mu-1)(\mu-1)}{\mu-2}\eta_{1}\,,\\
g_{\phi\phi O}^{2} & = &
\frac{2}{N}\frac{\Gamma(2\mu-2)}{\Gamma(3-\mu)\Gamma^{3}(\mu-1)}\Bigl(1
+\frac{1}{N}\,g_{1}\Bigl)\,,\nonumber \\
g_{1} & = &
-2\left(\frac{2\mu^{2}-3\mu+2}{\mu-2}{\cal{C}}(\mu)+\frac{8\mu^{3}-24\mu^{2}
+21\mu-2}{2(\mu-1)(\mu-2)}\right)\eta_{1},
\end{eqnarray}
while the results for $C_{T}$ and $C_{J}$ read
\begin{eqnarray}
C_{T} & = &
\frac{Nd}{(d-1)S_{d}^{2}}\left(1+\frac{1}{N}\,C_{T,1}\right),\,C_{T,1}
=  -\left(\frac{2}{\mu
+1}{\cal{C}}(\mu)+\frac{\mu^{2}+3\mu-2}{\mu(\mu-1)(\mu+1)}\right)
\eta_{1},\,\\
C_{J} & = &
\frac{2}{(d-2)S_{d}^{2}}\left(1+\frac{1}{N}\,C_{J,1}\right),\,C_{J,1}
=  -\frac{2(2\mu-1)}{\mu(\mu-1)}
\eta_{1},\,
\end{eqnarray}
with ${\cal{C}}(\mu) =\psi(3-\mu)+\psi(2\mu-1)-\psi(1)-\psi(\mu)$,
where $\psi(x)=\Gamma'(x)/\Gamma(x)$.

\section{Discussion of the Results}

The next-to-leading order in $1/N$ values of $\eta$ and
$\eta_{o}$ in (19) and (20) coincide with corresponding results in the
Lagrangian formulation of the $O(N)$ vector model e.g. see
\cite{Ruhl1} and references therein. The results (23) for $C_{J}$ and
(21) for $g_{\phi\phi O}$ were  first derived in
\cite{Ruhl2} and \cite{Ruhl3} correspondingly. An expression for the
next-to-leading order correction in
$1/N$ for $C_{T}$
 was given in \cite{Ruhl3} but it is different from our result
(22). We explain below why our result (22) is believed to be  correct.

Our results (19)-(23) give the values of the
dynamical parameters at the non-universal IR fixed point of the $O(N)$
vector model in  $2<d<4$ \footnote{The end
values $d=2,4$ require separate discussion. We intend to pursue such an
investigation in the future.}. It is believed that this IR fixed point is
connected through the  RG flow with an UV one, which corresponds to a
free (Gaussian)
theory of $N$ massless scalars. However, this RG flow is
nonperturbative over the parameter space since it cannot be seen in a
weak coupling expansion
but only in the context of the $1/N$ expansion. It is also worth noting
that the UV
Gaussian theory has a field of dimension $d-2$ while the IR
non-trivial theory has a field of dimension
$2+O(1/N)$. Therefore, for $N\rightarrow \infty$ and
e.g.  $d=3$ they
seem to be different theories since the have different field
content. However, the duality property discussed
in \cite{anastasios2} may relate these two theories at the
expense of unitarity.

In
Fig.3 we plot $C_{T,1}$ and $C_{J,1}$ for $2<d<4$ and by virtue of
(22) and (23) we see that, at least to the order in $1/N$ considered here,
$C_{T}$(UV)$> C_{T}$(IR) and $C_{J}$(UV)$> C_{J}$(IR). This is
in accord with a possible generalisation of the $C$- and $k$-theorems
in higher dimensions requiring $C_{T}$ and $C_{J}$ to be the
fixed point values of the $C$- and $k$-functions respectively.
\begin{figure}[t]
\epsfxsize=10cm
\epsfbox{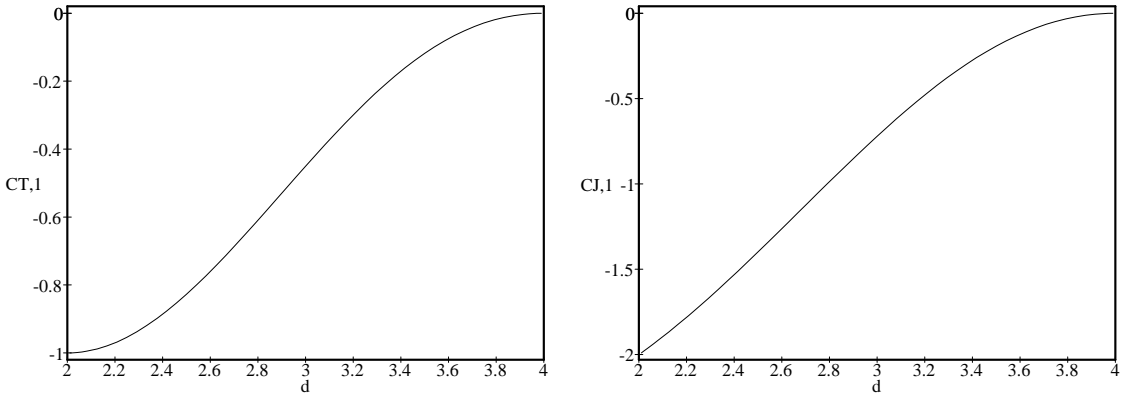}
\caption{$C_{T,1}$ and $C_{J,1}$ for $2<d<4$.}
\end{figure}

In two dimensions the $C$-function counts, properly normalised,
the massless degrees of freedom in the theory on a particular length
scale \cite{Cardy1}. Adopting such an interpretation for the
$C$-function in higher
dimensions we may normalise $C_{T}$ in (16) to $N$  e.g. the number of
massless scalar fields. It is well known \cite{Polyakov} that the
existence of a
non-trivial IR fixed point in the $O(N)$ vector model
for $2<d<4$ is an indication for the symmetry breaking pattern
$O(N)\rightarrow O(N-1)$ in the theory.  That is, the IR critical
point separates the $O(N)$
symmetric phase  having  $N$ massive  modes, from the $O(N-1)$ symmetric
one with $N-1$ massless Goldstone bosons and one
massive mode. Hence, at this fixed point the potential $C$-function,
properly normalised, should have $N-1$ as lower bound. Indeed, for
$2<d<4$ we see from the first graph in Fig.3 that $N-1<C_{T}$(IR) is
satisfied which is another test for the validity of our result (22).

The $k$-function on the other hand is connected with the internal
symmetry of the theory i.e. in a theory with no such symmetry a
$k$-function does not exist. If an internal symmetry remains unbroken
along the RG flow from the UV to the IR fixed point we expect that
$k$(UV)$=k$(IR). We may therefore interpret the $k$-function as
counting the {\it{amount of internal symmetry}} in the theory on a
particular length scale. At the fixed points the number
of conserved currents, i.e. $N(N-1)/2$ for a $O(N)$ invariant
theory, may be taken as   a quantity which counts
the {\it{amount of internal symmetry}}. Hence, we expect that the
$k$-function at the IR fixed
point of the $O(N)$ vector model for $2<d<4$, normalised to 1 in the
$O(N)$ symmetric phase, should satisfy $1-(2/N)<k$(IR). Indeed,
from (16), (23) and the second graph in Fig. 3 we see that
$1-(2/N)<C_{J}$(IR) is remarkably satisfied.

Another crucial test for our results (22) and (23) is to compare them with
known results in
the context of $\e$-expansion when $\e=4-d>0$. In four dimensions, if the
theory is
defined for a background metric $g_{\mu\nu}$ and has a gauge field
$A_{\mu}^{\a}$ coupled to the conserved vector current, even for a
conformal theory there is a trace anomaly \cite{anastasios1,Birrel}
\begin{equation}
g^{\mu\nu}\la
T_{\mu\nu}\ra=-\b_{a}F-\kappa{\textstyle{\frac{1}{4}}}
F_{\mu\nu}^{\a}F^{\a,\mu\nu}+\cdots,
\end{equation}
where $F$ is the square of the Weyl tensor and terms which are
irrelevant here are neglected. The quantities $\b_{a}$
and $\kappa$ can be perturbatively calculated and for a  $O(N)$
invariant
renormalisable  field
theory with ${\textstyle{\frac{1}{24}}}g\left(\phi^{2}\right)^{2}$
interaction, a three-loop
calculation yields \cite{Jack}
\begin{eqnarray}
\b_{a} & =  &
-\frac{1}{16\pi^{2}}\frac{N}{120}\Bigl(1-\frac{5}{108}
(N+2)u^{2}\Bigl), \\
\kappa & = &
\frac{1}{3}\frac{1}{16\pi^{2}}\,R\,\Bigl(1-\frac{1}{12}
(N+2)u^{2}\Bigl),
\end{eqnarray}
where $u=g/16\pi^{2}$ with $g$ the renormalised coupling
and
$\mbox{tr}(t^{\a}t^{\b})=-\delta^{\a\b}R$. For
the adjoint representation of $O(N)$ we have
$(t^{\a\b})^{\c\delta}=-(\delta^{\a\c}\delta^{\b\delta}
-\delta^{\a\delta}\delta^{\b\c})$
and $R=2$. The results of our previous work \cite{anastasios1} show
that   for a conformal theory when  $d=4$
\begin{equation}
C_{T}=-\frac{640}{\pi^{2}}\b_{a}\,\,\,\,\,,\,\,\,\,\,C_{J}
=\frac{6}{\pi^{2}}\kappa.
\end{equation}
In general we suppose that we may
write $C_{T}(\e,u_{*})$, $C_{J}(\e,u_{*})$ where $u_{*}$ is the
critical coupling. The free or Gaussian field theory results
in (16) correspond to $C_{T,f}=C_{T}(\e,0)$ and
$C_{J,f}=C_{J}(\e,0)$
while (22) and (23)  give  $C_{T}(0,u)$ and $C_{J}(0,u)$. Using then
(25) and (26)  with $u_{*}=3\e/(N+8)+O(\e^{2})$
\cite{Zinn-Justin} gives the leading corrections in the $\e$-expansion
\footnote{The result (28) for $N=1$ was found in \cite{Cappelli}.}
\begin{eqnarray}
C_{T} & = &
C_{T,f}\Bigl(1-\frac{5}{12}\frac{N+2}{(N+8)^{2}}\e^{2}
+O(\e^{3})\Bigl),
\\
C_{J} & = &
C_{J,f}\Bigl(1-\frac{3}{4}\frac{N+2}{(N+8)^{2}}\e^{2}
+O(\e^{3})\Bigl).
\end{eqnarray}
As $d\rightarrow 4$ we see from (19) that
$\eta_{1}\sim\e^{2}/4$ and then we can easily show  that
out results (22) and (23)  agree correspondingly with
(28)
and (29), something  which is a remarkable independent
check for their validity at least up to the order considered here.

Finally, we note that it was shown in  \cite{Cardy2}  that $C_{T}$
parametrises universal finite size effects of statistical systems at their
critical points in two dimensions which provides a natural method for
its measurement both numerically and experimentally. For $d>2$,
although  Cardy
\cite{Cardy2} has pointed out that $C_{T}$ may be in principle
measurable, the finite scaling of the
free energy is parametrised \cite{Fradkin} by a universal number
$\tilde{c}$ whose relation with $C_{T}$ is not  clear. Sachdev
\cite{Sachdev} has calculated $\tilde{c}$ for the $O(N)$ vector model
to leading order in $1/N$ for $d=3$ and found it to be a rational number
however
different from the leading order in $1/N$ value for $C_{T}$
\footnote{It is interesting to point out  that the leading order in
$1/N$  value
for $C_{T}$
coincides with the Gaussian theory value $C_{T,f}$ in any dimension
whereas as shown in \cite{Sachdev} the leading order in $1/N$ and the
Gaussian values for $\tilde{c}$ differ in $2<d<4$.}. Using our results
(22) and (23) we obtain for $d=3$
\begin{eqnarray}
C_{T}|_{d=3} & = &
N\frac{3}{2S_{3}^{2}}\Bigl(1-\frac{1}{N}\frac{40}{9\pi^{2}}\Bigl),
\\
C_{J}|_{d=3} & = &
\frac{2}{S_{3}^{2}}\Bigl(1-\frac{1}{N}\frac{32}{9\pi^{2}}\Bigl).
\end{eqnarray}
With our normalisation
$C_{T}$
has to be multiplied by $S_{d}^{2}/2$ to agree with  the corresponding
quantity in  \cite{Sachdev}, and then we can answer by virtue of (30) part
of the question addressed in that reference: $C_{T}$ does not seem to be a
rational number for finite $N$ in three dimensions.

\vspace{2cm}
{\bf{Acknowledgments}}
\small

I am indebted to Professor John Cardy for a very illuminating discussion.

\end{document}